\documentclass[11pt,twoside]{article}

\usepackage{asp2006}
\usepackage{epsf}
\usepackage{psfig}
\usepackage{lscape}

\begin{document}
\title{Spiral and Bar Instabilities Provoked by Dark Matter Satellites}   
\author{John Dubinski\altaffilmark{1}, Jean-Ren\'e Gauthier\altaffilmark{1,2}, Larry
Widrow\altaffilmark{3}, and Sarah Nickerson\altaffilmark{1}}   
\altaffiltext{1}{Department of Astronomy and Astrophysics, University of Toronto, Toronto, ON M4S 3H4,
Canada dubinski@astro.utoronto.ca; nickerson.sarah@gmail.com}    
\altaffiltext{2}{Department of Astronomy and Astrophysics, Kavli Institute for Cosmological Physics, The University of Chicago, Chicago, IL 60637, USA gauthier@oddjob.uchicago.edu}    
\altaffiltext{3}{Department of Physics, Queen's University, Kingston, ON K7L 3N6, Canada widrow@astro.queensu.ca} 

\begin{abstract} 
We explore the secular dynamical evolution of an N-body model 
of M31 in the
presence of a population of 100 dark matter satellites over 10 Gyr.  
The satellite
population has structural and kinematic characteristics
modelled to follow the predictions of $\Lambda$CDM cosmological simulations.
Vertical disk heating is a small effect despite many interactions
with the satellite population with only a 20\% increase in vertical velocity
dispersion $\sigma_z$ and the disk scale height $z_d$ at the equivalent
solar radius $R=2.5 R_d$.   However, the stellar disk is noticeably flared after
10 Gyr with $z_d$ nearly doubling at the disk edge.  Azimuthal disk heating
is much larger with $\sigma_R$ and $\sigma_z$ both increasing by $1.7\times$. However, in a
control experiment without satellites dispersion increases by $1.5\times$ suggesting that
most of the effect is due to heating through scattering off of spiral structure
excited by swing-amplified noise.  Surprisingly, direct impacts of satellites on
the disk can excite spiral structure with a significant amplitude and in some cases
impacts close to the disk center also induce the bar instability.  
The large number of dark
matter satellite impacts expected over a galaxy's lifetime may be a significant
source of external perturbations for driving disk secular evolution.
\end{abstract}


\section{Introduction}

Cosmological simulations in $\Lambda$CDM show that 
the dark matter halos of galaxies contain
hundreds to thousands of subhalos or dark satellites \citep{moo99,gao04,die07}.  The existence of subhalos
raises interesting questions about the
dynamical evolution of disk galaxies.  While 
this population is under-represented in the observed satellites
\citep{kly99}, the recent success of
$\Lambda$CDM in accounting for many properties of the
universe from the CMB to the large-scale structure leads us to take this
prediction seriously. Cosmological infall of a large number of dark
satellites onto a typical spiral galaxy could be a major source of disk heating and
thickening \citep{tot92,fon01,ben04,kaz07} beyond known astrophysical processes 
so it important to quantify the effect and see if the observed galaxy population is
morphologically consistent.
A main difficulty with
numerical studies is that N-body disks are prone to self-heating by two-body
relaxation with inadequate resolution so that any perturbations by
satellites can easily be masked by numerical effects.  Ad hoc assumptions about
the structure of satellites also make the interpretation of results difficult in
light of the predictions of $\Lambda$CDM.

In this study, we attempt to overcome these problems using new galaxy simulations with
sufficient resolution to measure the heating directly due to a 
cosmologically inspired model population of dark matter satellites.  These simulations
reveal the importance of satellite impacts -- i.e direct passage of satellites
through the stellar disk -- in exciting both spiral and bar instabilities.  Dark
satellite interactions may therefore be an essential driver of secular dynamical evolution
of disk galaxies (see also \citet{kaz07}). 

\section{Methods}

For the study here, we use the stable, equilibrium  axisymmetric model 
of M31 derived from a distribution function as described in
\citet{wid05}.  We model M31 with an exponential disk, a Hernquist model bulge,
and a cuspy dark halo with an NFW profile.  Model parameters are determined 
that fit the rotation curve
and surface brightness profile for M31 with an assumed $M/L$ ratio for the stellar
components such that the disk remains
stable against bar formation for 10 Gyr.

The method for generating a
satellite population 
is described in detail in
\citet{gau06}.  The satellite properties
reflect cosmological numerical predictions for 
the subhalo mass function, radial distribution, tidal
radii as well as internal density  structure.
To summarize, 10\% of the dark matter halo mass is initially in 100
subhalos spanning a mass range of $10^8 - 10^{10}$ M$_\odot$ selected from a mass
function $dN/dM \sim M^{-1.8}$.  The radial number density of satellites are set
according to the formulae presented in \citet{gao04}.  Our highest
resolution galaxy model contains the following numbers of particles: 10M disk, 5M
bulge,  and 20M smooth halo for a total of 35M.  
The 100 dark satellites each contain 100K particles
for a total of 10M.  We run the simulation using a parallelized treecode
\citep{dub96} with a fixed Plummer softening length $\epsilon=15$ pc and 20000
equal timesteps with $\delta t=0.49$ Myr.  Total binding energy is conserved to
within 0.3\% and total angular momentum is conserved to within 2\%.  We also use
models with 10 times fewer particles in some additional studies.

\section{Disk Heating}
We first ran a control simulation at 35M particles to understand numerical
effects (Fig. 1).  Vertical disk heating is negligible over most of the disk with
almost no change in $\sigma_z$ and the vertical scale height at this resolution.  
However, the radial and
azimuthal velocity dispersions grow by 50\% due to spiral instabilities arising
from swing amplified Poisson noise in the disk particle distribution.
We then added satellites in two different runs with statistically
similar distributions.  In our first simulation, the disk developed
a bar half way through the run \citep{gau06} and so introduces an unwanted
additional source of heating (see Movie 1).  The formation of a bar was
unexpected and we will discuss the origin of this instability shortly.
Fortunately in the second run, no bar formed so we could directly measure heating
effects by the satellite population (Fig. 1).  Vertical
heating is still small with roughly a 20\% increase in $\sigma_z$ even under
the bombardment of satellites.  The increase in $\sigma_r$ and $\sigma_\theta$ is also
only 20\% over and above the heating caused by intrinsic spiral instabilities.
However, there is a noticeable flaring of the disk in the presence of satellites
with the scale height nearly doubling from 2 disk scale lengths to the disk edge.
\citet{kaz07} also see this effect in similar work.  Similar features are seen in the
outer disk of M31 \citep{iba07}.

\begin{figure}
\plotone{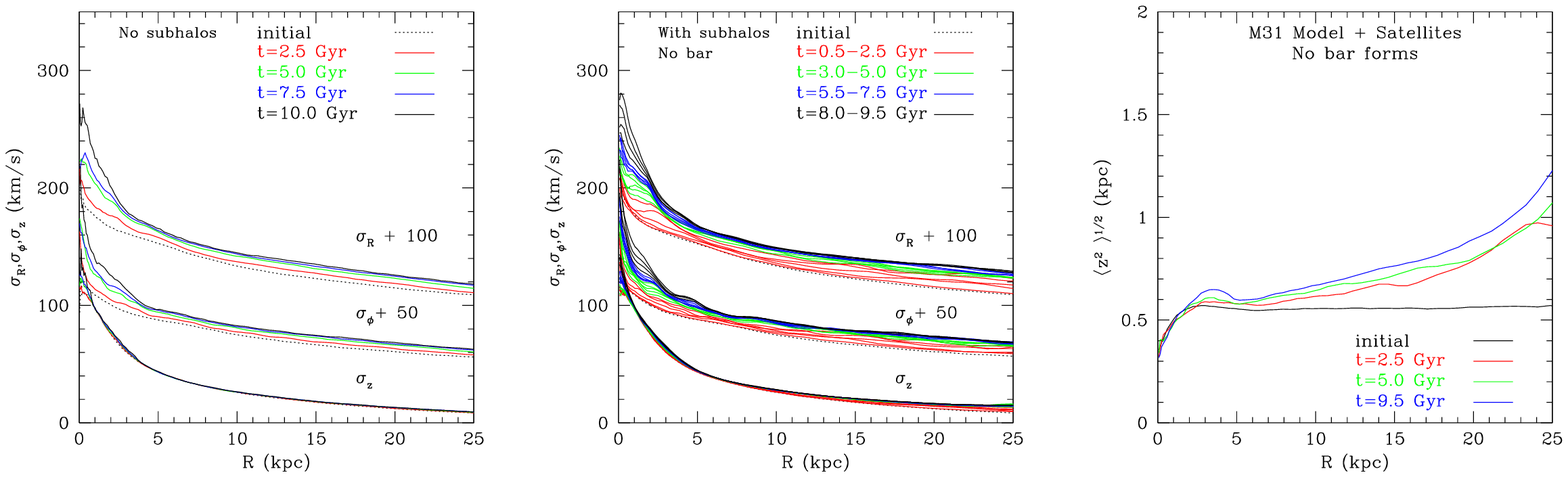}
\caption{The evolution of the disk velocity ellipsoid for the control simulation (no
subhalos) and the simulation with 100 dark satellites with no bar instability.
Also, the evolution of the vertical scale-height as measured by the variance in
$z$ at different radii. 
}
\end{figure}


\section{Spiral and Bar Instabilities from Satellite Impacts}
An unexpected feature of these simulations is the induced bar instability in our first
run as well
as easily distinguished multi-armed global spiral structure.  The main cause of
these features are satellite impacts.  The passage of a satellite
through the disk induces a localized disturbance that presumably grows by
Toomre's swing amplification mechanism \citep{too81}.  Note that the tidal
effects of the satellites are generally small and so this mechanism
is quite different than the tidal interactions
responsible for grand-design spiral galaxies like M51.  The mechanism is closer to
the original ideas suggested by both \citet{gol65} and \citet{jul66} where a mass
perturbation appearing within the disk -- a giant molecular cloud or 
massive star forming region -- is the source of a disturbance that is subsequently
amplified.

A virtual fly by of the evolving galaxy clearly shows episodes of spiral
structure excitation immediately after satellites pass through the disk (see Movie
2).  In the case of model with the bar instability, there appears to be a single
strong encounter with one of the more massive satellites near the center of disk
just before the onset of the bar.   It seems likely the disturbance caused by the
passing satellite disrupts the center of the galaxy enough to make it susceptible
to the swing-amplifier feedback loop \citep{too81}.   

We are testing these ideas further with more experiments and
idealized perturbations representing satellite passages throught the disk.  We performed 10 additional experiments with 4.5M
particle models and statistically independent but consistent satellite populations
in orbit around M31 (see Movie 3).  Five out of ten of these models develop a
bar instability during their lifetime apparently due to chance central encounter
with a massive satellite.  Those models that do not suffer strong central encounters with
satellites avoid bar formation.  In all cases, spiral perturbations of
significant amplitude are observed due to satellite interactions.  

We are now doing a quantitative study of the effect of satellite impacts on disks
using a transient mass perturbation appearing within the plane of the disk.
Preliminary results suggest that even satellite
masses as small as $10^9$ M$_\odot$ or roughly 5-10\% the mass of the LMC are
large enough to induce obvious spiral structure while the passage of a LMC-sized satellite creates a
strong response (see Movie 4).
A typical galaxy will experience
dozens of impacts with satellites more massive than $10^9$ M$_\odot$ during its life according
to the predictions of $\Lambda$CDM.  Dark satellite impacts may then play an important
role in maintaining the multi-armed, global spiral patterns seen in the disk galaxies.
At this stage, we need to quantify the response of a disk as a function of satellite
impactor mass and impact radius on the disk.  The orbital statistics of $\Lambda$CDM
subhalos from cosmological simulations will allow us to determine the frequency, mass
distribution and distribution of impact radii expected on a galactic disk over its
history.  By combining these two results, we should be able to quantify the effect of
dark satellites on disk secular evolution and so address observations of the
morphological appearance of galaxies including the bar fraction \citep{jog04} throughout
cosmic history.

\acknowledgements 
The authors acknowledge supercomputing time on machines supported by SHARCNET and CITA.
We also acknowledge funding by NSERC.

\noindent {\bf MOVIES 1,2,3 \& 4} are available at: \\ 
http://www.cita.utoronto.ca/$\sim$dubinski/Rome2007



\begin{thebibliography}{}
\bibitem[Benson et al.(2004)]{ben04} Benson, A.J., Lacey, C.G., Frenk, C., Baugh, C.M., Cole, S., 2004,
\mnras, 351, 1215
\bibitem[Diemand et al.(2007)]{die07} Diemand, J., Kuhlen, 
M., \& Madau, P.\ 2007, \apj, 667, 859
\bibitem[Dubinski(1996)]{dub96} Dubinski, J.\ 1996, New 
Astronomy, 1, 133 
\bibitem[Font et al.(2001)]{fon01} Font, A.S., Navarro, J.F., Stadel, J., Quinn, T., 2001, \apj, 563, L1
\bibitem[Gao et al.(2004)]{gao04} Gao, L., White, S.D.M., Jenkins, A., Stoehr, F., Springel, V. 2004, \mnras, 355, 819
\bibitem[Gauthier et al.(2006)]{gau06} Gauthier, J.-R., 
Dubinski, J., \& Widrow, L.~M.\ 2006, \apj, 653, 1180 
\bibitem[Goldreich \& Lynden-Bell(1965)]{gol65} Goldreich, 
P., \& Lynden-Bell, D.\ 1965, \mnras, 130, 125 
\bibitem[Ibata et al. (2007)]{iba07} Ibata, R., Martin, N.F., Irwin, M., Chapman, S.,
Ferguson, A.M.N., Lewis, G.F., \& McConnachie, A.W. 2007, preprint,
http://arxiv.org/abs/0704.1318
\bibitem[Julian \& Toomre(1966)]{jul66} Julian, W.~H., \& 
Toomre, A.\ 1966, \apj, 146, 810 
\bibitem[Jogee et al.(2004)]{jog04} Jogee, S., et al.\ 2004, 
\apjl, 615, L105 
\bibitem[Kazantzidis et al.(2007)]{kaz07} Kazantzidis, S., Bullock, J.S., Zenter, A.R.,
Kravtsov, A.V., \& Moustakas, L.A. 2007, preprint, http://xxx.lanl.gov/abs/0708.1949 
\bibitem[Klypin et al.(1999)]{kly99} Klypin, A., Kravtsov, A.V., Valenzuela, O., 1999, \apj, 522, 82
\bibitem[Kuijken \& Dubinski(1995)]{kui95} Kuijken, K., Dubinski, J. 1995 \mnras, 277, 1341
\bibitem[Moore et al.(1999)]{moo99} Moore, B., Ghigna, S., Governato, F., 1999, \apj, 524, L19
\bibitem[Toomre(1981)]{too81} Toomre, A.\ 1981, Structure and 
Evolution of Normal Galaxies, 111 
\bibitem[T\'oth \& Ostriker(1992)]{tot92} T\'oth G., Ostriker, J.P., 1992, \apj, 389, 5
\bibitem[Widrow \& Dubinski(2005)]{wid05} Widrow, L.~M., \& 
Dubinski, J.\ 2005, \apj, 631, 838 
\end{thebibliography}
\end{document}